\def\kms{\,{\rm km}\,{\rm s}^{-1}}
\def\hmpc{\,{h^{-1} {\rm Mpc}}}
\def\mpch{\,{h {\rm Mpc}^{-1}}}
\def\fnl {\,f^{\rm{local}}_{\rm{NL}}}
\def\fnlabs {\,\left|f^{\rm{local}}_{\rm{NL}}\right|}
\def\msun {{\rm M}_{\odot}}

\documentclass[usenatbib]{mn2e}
\usepackage{epsfig}

\usepackage{amssymb}
\usepackage{amsmath}
\usepackage{aecompl}
\usepackage[T1]{fontenc}

\begin{document}

\title[Primordial non-Gaussianity and velocity field]{%
Independent constraints on local non-Gaussianity from the peculiar
velocity and density fields}

\author[Y.-Z.~Ma, J.~E.~Taylor and D.~Scott]{Yin-Zhe Ma$^{1,2,\dagger}$, James E. Taylor$^{3,\ddagger}$, \& Douglas Scott$^{1,\star}$\\
$^1$Department of Physics and Astronomy, University of British Columbia, Vancouver, V6T 1Z1, BC Canada.\\
$^2$Canadian Institute for Theoretical Astrophysics, Toronto, Canada.\\
$^3$ Department of Physics and Astronomy, University of Waterloo, Waterloo, Ontario, Canada N2L 3G1. \\
emails: $^{\dagger}$mayinzhe@phas.ubc.ca;\,
$^{\ddagger}$taylor@uwaterloo.ca;\, $^{\star}$dscott@phas.ubc.ca}


\maketitle

\begin{abstract}

Primordial, non-Gaussian perturbations can generate
scale-dependent bias in the galaxy distribution. This in turn will
modify correlations between galaxy positions and peculiar
velocities at late times, since peculiar velocities reflect the
underlying matter distribution, whereas galaxies are a biased
tracer of the same. We study this effect, and show that
non-Gaussianity can be constrained by comparing the observed
peculiar velocity field to a model velocity field reconstructed
from the galaxy density field assuming linear bias. The amplitude
of the spatial correlations in the residual map obtained after
subtracting one velocity field from the other is directly
proportional to the strength of the primordial non-Gaussianity. We
construct the corresponding likelihood function and use it to
constrain the amplitude of the linear flow $\beta$ and the
amplitude of local non-Gaussianity $\fnl$. Applying our method to
two observational data sets, the Type-Ia supernovae (A1SN) and
Spiral Field \textit{I}-band (SFI++) catalogues, we obtain
constraints on the linear flow parameter consistent with the
values derived previously assuming Gaussianity. The marginalised
1-D distribution of $\fnlabs$ does not show strong evidence for
non-zero $\fnl$, and we set $95\%$ upper limits $\fnlabs<51.4$
from A1SN and $\fnlabs<92.6$ from SFI++. These limits on $\fnl$
are as tight as any set by previous large-scale structure
measurements. Our method can be applied to any survey with radial
velocities and density field data, and provides an independent
check of recent CMB constraints on $\fnl$, extending these to
smaller spatial scales.

\end{abstract}


\section{Introduction}

\label{vel_intro}

In the standard cosmological model, the large-scale structure of the universe
has its origin in quantum fluctuations generated during inflation.
The simplest single-field, slow-roll inflation model is predicted to generate primordial
scalar perturbations that are close to Gaussian and scale-invariant %
\citep{Bardeen86}. There exist, however, a large class of
alternative models of inflation that can generate significant
non-Gaussian components of the gravitational potential \citep[e.g.][]{Wands10}.
The degree of non-Gaussianity is usually expressed as the amplitude of the
bispectrum $B$ normalized by the power spectrum $P$, that is
\begin{equation}
f_{\mathrm{NL}} \equiv \frac{B(k_1,k_2,k_3)}{2\left[P(k_1)P(k_2)+P(k_2)P(k_3)+P(k_1)P(k_3)\right]},
\label{eq:fnldef}
\end{equation}
where $k_{1},k_{2}$ and $k_{3}$ are three $k-$space modes.
If one assumes that the final evolved potential $\Phi$ is a local function
of a primordial Gaussian field $\phi$, then the final potential can be approximated to second order as
\begin{equation}
\Phi = \phi + f_{\mathrm{NL}}^{\mathrm{local}}(\phi^2-\langle\phi^2\rangle).
\end{equation}
 This corresponds to the ``squeezed'' limit ($k_1^2 \ll k_2^2 + k_3^2$) of the triangle configuration.
 In this specific case, non-Gaussianity is scale independent, but more generally $f_{\mathrm{NL}}$ as defined in Eq.~(\ref{eq:fnldef})
 could depend on scale and on the configuration of the triangle.

The large-scale clustering of galaxies provides an important
observational test of primordial non-Gaussianity. Galaxies should
trace fluctuations in the underlying matter field to some degree,
but may be more or less clustered than the matter distribution as
a whole. If, for instance, galaxies form only in the densest peaks
of the matter field, which collapse into bound dark matter haloes,
these peaks will cluster more strongly than the field on average.
In the limit of small amplitude fluctuations, fluctuations in the
galaxy distribution $\delta _{\mathrm{g}}$ and fluctuations in the
matter distribution $\delta _{\mathrm{m}}$ can be assumed to be
proportional and related by a (linear) bias parameter  $b$:
$\delta _{\mathrm{g}}=b\delta _{\mathrm{m}}$. For peaks in a
Gaussian random field, the properties of this halo bias are well
understood \citep{Bardeen86}. In the case of (local)
non-Gaussianity, however, \citealt{Dalal08} and \citealt{Matt08}
have shown that the halo distribution is affected by an
additional, scale-dependent bias factor. In particular, a large
value of $f_{\mathrm{NL}}^{\mathrm{local}}$ implies that the
amplitude of the two-point correlation function is larger on large
scales than would be expected in the Gaussian case. A definitive
detection of this excess clustering signal could rule out the
single-field slow-roll inflation models, and yield insight into
the mechanism that drove the inflaton field in the early Universe.

Several studies have used observed clustering to constrain
non-Gaussianity. \cite{Xia11} found $\fnl=48 \pm 20$ [$1\sigma$
confidence level (CL)], using radio sources from the NRAO VLA Sky
Survey (NVSS), the quasar and MegaZ-LRG (DR7) catalogues of the
Sloan Digital Sky Survey (SDSS, \citealt{York00}), and the final
SDSS II Luminous Red Galaxy (LRG) photometric redshift survey.
\cite{Nik13} found $f_{\mathrm{NL}}^{\mathrm{local}} = 90\pm30$ at
$1\sigma$ CL using photometric SDSS data , but suggested that this
result may be better interpreted as
$f_{\mathrm{NL}}^{\mathrm{local}} < 120$ at $84\%$ CL, due to the
concern over systematics. In addition, \cite{Ross13} used SDSS-III
Baryon Oscillation Spectroscopic Survey (BOSS) data, which were
included in the SDSS data release nine (DR9) to
constrain the $f_{\mathrm{NL}}^{\mathrm{local}}$ value, and found $-45<f^{%
\mathrm{local}}_{\mathrm{NL}}<195$ at $2\sigma$ CL, and $P(f^{\mathrm{local}}_{%
\mathrm{NL}}>0)=91\%$.

Currently, however, the most stringent constraints on non-Gaussianity, at least in
a scale-independent form, come from measurements of the bispectrum of the Cosmic
Microwave Background (CMB).
In 2011, the 7-year Wilkinson Microwave Background Probe
(\textit{WMAP}) data were used to obtain $-10
< f_{\mathrm{NL}}^{\mathrm{local}} < 74$ at $95\%$ CL %
\citep{Komatsu11}. Later, the 9-year \textit{WMAP} data
\citep{Bennett12} were used to provide a similar constraint $-3 <
f_{\mathrm{NL}}^{\mathrm{local}} < 77$ at $95\%$ CL. Most
recently, \textit{Planck} released its nominal mission survey
results, which constrain local non-Gaussanity
to be $f_{\mathrm{NL}}^{\mathrm{local}}=2.7 \pm 5.8$ at $68\%$ CL %
\citep{Planck24}. Although these constraints have
already placed tight limits on many variant models of inflation, they only probe
non-Gaussianity over a limited range of scales and geometries.
Thus in principle it remains interesting to develop complementary tests of
non-Gaussianity based on large-scale structure.
In this paper we introduce a new method for constraining primordial
non-Gaussianity by using measurements of the local peculiar
velocity field.

In the standard gravitational instability picture, the
peculiar velocity field is induced by the gravitational pull of inhomogeneities
in the matter distribution, and can be expressed as ${\rm \pi}$ %
\citep{Peebles80}
\begin{equation}
\mathbf{v}_{\mathrm{g}}(\mathbf{r})=\frac{H_{0}f_{0}}{4 \pi}\int d^{3}%
\mathbf{r}^{\prime }\delta _{\mathrm{m}}(\mathbf{r}^{\prime },t_{0})\frac{(%
\mathbf{r}^{\prime }-\mathbf{r})}{\left\vert \mathbf{r}^{\prime }-\mathbf{r}%
\right\vert ^{3}},  \label{eq:vg0}
\end{equation}%
where $H_{0}=H(t_{0})$ is the Hubble parameter at the present epoch,
$f_{0}$ is the present day growth rate (henceforth we drop the subscript $0$) and $%
\delta _{\mathrm{m}}$ is the underlying dark matter perturbation, i.e. $%
\delta _{\mathrm{m}} \equiv (\rho -\overline{\rho
})/\overline{\rho }$. Assuming galaxies roughly trace
the underlying matter distribution on large scales, then the
density contrasts of the two should be related by a linear, deterministic bias factor,
$\delta _{\mathrm{g}}=b\delta _{\mathrm{m}%
}$. On the other hand, if primordial non-Gaussianity exists, the
relationship between $\delta _{\mathrm{g}}$ and $\delta
_{\mathrm{m}}$ is no longer a single constant bias factor,
but depends on scale. This scale-dependent bias may in turn cause a
scale-dependence in the relationship between the velocity and density fields.
A measurement of this relationship would therefore constrain primordial non-Gaussianity.

Before we move on, we should mention that if $\delta_{\rm g}$ and $\delta_{\rm m}$
are related by a constant bias factor, one can replace the growth rate of density fluctuations
$f$ with the dimensionless ``linear flow" parameter $\beta \equiv f/b$ (e.g. \citealt{Ma12a}).
The amplitude of the peculiar velocity field scales linearly with $\beta$; its value can be estimated
by comparing peculiar velocities derived from distances and redshifts
with a model velocity field reconstructed from the density distribution
using Eq.~(\ref{eq:vg0}). The estimated value for the linear flow parameter is $\beta \simeq 0.54$ \citep{Ma12a}.
Many previous studies have performed this ``$v$-$v$'' analysis, comparing the observed and reconstructed velocity fields;
the two match each other fairly well, which constitutes good observational evidence for the gravitational instability paradigm %
\citep{Davis96,Branchini01,Scoccimarro01,Feldman01,Verde02,Ma12a}.
In this paper, we want to extend this method to include
the contribution from primordial non-Gaussianity, and use
state-of-art peculiar velocity field data sets to constrain
$f^{\mathrm{local}}_{\mathrm{NL}}$.

This paper is organized as follows. In Sect.~\ref{sec:model}, we
will model the non-Gaussianity, establish its relation to the
measured and reconstructed velocity fields, and construct the
likelihood function. In Sect.~\ref{sec:data}, we present the
observed and modelled peculiar velocity data, which we will use to
constrain the primordial non-Gaussianity. We then present and
discuss the results of the likelihood analysis in
Sect.~\ref{sec:results}, and compare the models with and without
non-Gaussianity. The conclusions will be presented in the last
section.

Throughout the paper, we assume a spatially flat cosmology with \textit{%
Planck} parameter values \citep{Planck16}, i.e. fractional matter density $%
\Omega_{\rm m} = 0.3183$, fractional baryon density multiplied by
Hubble constant squared $\Omega_{\rm b}h^{2}=0.022$, fractional
cold dark matter density with Hubble constant squared $\Omega_{\rm
c}h^{2}=0.12038$, Hubble constant $h=0.67$ (in unit of $100 \,
{\rm km}\, {\rm s^{-1}} \, {\rm Mpc^{-1}}$), spectral index of
primordial power spectrum $n_{\rm s}=0.962$, and amplitude of
fluctuations $\sigma_8=0.83$.

\section{Method}

\label{sec:model}

In this section, we first discuss the physics of the density and
velocity fields of galaxies (Sect.~\ref{sec:physics}), and then
construct a likelihood method to quantify the non-Gaussianity
present in the local density field (Sect.~\ref{sec:statistics}).

\subsection{Scale-dependent bias from non-Gaussianity}
\label{sec:physics}

As shown in both \cite{Materrese00} and later in \cite{Dalal08},
\cite{Matt08} and \cite{Wands09}, primordial non-Gaussianity
changes the mass function of dark matter haloes, with positive
$f^{\mathrm{local}}_{\mathrm{NL}}$ increasing the abundance of
high-mass haloes. Non-Gaussianity also changes the bias of the
halo distribution relative to the matter distribution, however.

In the Gaussian case, this bias has two components
\citep{Dalal08}: the Lagrangian bias ($b_{\rm L}=b-1$), which
reflects the contribution of long-wavelength modes to boosting
peaks over the threshold for collapse, and the Eulerian bias, an
extra factor of unity which reflects the net excess of matter in
over dense regions, or equivalently the motions of primordial
peaks at later times. Primordial non-Gaussianity affects the
initial conditions (which peaks in the primordial density field
become haloes), but not the subsequent gravitational evolution as
peaks get advected along with bulk matter flows. Thus the quantity
that determines the non-Gaussian bias correction is the Lagrangian
bias $b_{\rm L}$, not the Eulerian bias $b$.

\begin{figure}
\centerline{\includegraphics[bb=0 0 543
353,width=3.3in]{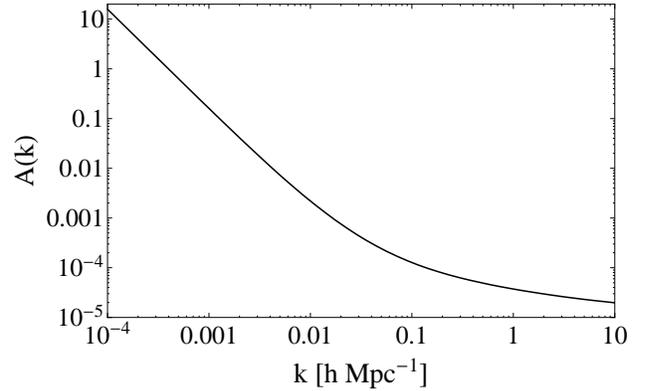}} \caption{Scale-dependence $A(k)$ of the non-Gaussian bias factor $\Delta b(k)$
(cf.~Eq.~(\ref{eq:Ak})).}
\label{fig:ak}
\end{figure}

The non-Gaussian contribution causes a scale-dependent bias in the
power spectrum of dark matter haloes. If one expresses the
(Gaussian) dark matter halo bias as a constant factor $b$, the
total additional bias is \citep{Dalal08,Matt08,Ross13}
\begin{equation}
\Delta b(k) = (b-1)f_{\mathrm{NL}}^{\mathrm{local}}A(k),
\label{eq:Bfnl}
\end{equation}
where the function $A(k)$ is
\begin{equation}
A(k) = \frac{3\delta_{\rm c}(z)\Omega_{\rm m} h^{2}}{k^2T(k)}\left(\frac{H_0}{c}%
\right)^2.  \label{eq:Ak}
\end{equation}
Here $k$ is in units of $h\,\text{Mpc}^{-1}$, $T(k)$ is the
transfer function, and $\delta_{\rm c}(z) = 1.686/D(z)$ is the
critical over-density for dark matter haloes to collapse at
redshift $z$ from the spherical collapse model, while $D(z)$ is
normalized to unity at $z=0$ \citep{Ross13}. In our case, since
most samples are within $100\hmpc$, we take $D(z=0)=1$ in our
analysis. We calculate the transfer function and linear matter
power spectrum $P_{\mathrm{m}}(k)$ by using the public software
package {\sc camb} \footnote{%
http://camb.info/} \citep{Lewis00}. We plot the function $A(k)$ in
Fig.~\ref{fig:ak}. One can see that the effect of local
non-Gaussianity is to increase bias preferentially on large
scales. If there is large local non-Gaussianity, it means that
there is a correlation between short- and long-wavelength modes,
since the geometry of local non-Gaussianity is a triangle with two
very long $k$-vectors and one short one ($k_{1}\simeq k_{2} \gg
k_{3}$). This means that the locations of massive haloes will be
correlated (or anti-correlated) with peaks in the very long
wavelength modes of the matter distribution. The function $A(k)$
describes how much enhancement is obtained from this correlation
on different scales.

The total bias is a combination of Gaussian linear bias and the additional non-Gaussian
bias (Eq.~(\ref{eq:Bfnl})):
\begin{equation}
b_{\mathrm{tot}} = b + \Delta b(k).  \label{eq:btot}
\end{equation}
Then because $\delta_{\rm g}=b_{\rm{tot}}(k) \delta_{\rm m}$, the
galaxy power spectrum $P_{\mathrm{g}}(k)$ is related to the
underlying matter power spectrum through
\begin{eqnarray}
P_{\mathrm{g}}(k) &=& b^{2}_{\mathrm{tot}}(k)P_{\mathrm{m}}(k)  \notag \\
&=& b^2P_{\mathrm{m}}(k) +2b\Delta b(k)P_{\mathrm{m}}(k) + \Delta
b^{2}(k) P_{\mathrm{m}}(k).  \label{eq:Pexp}
\end{eqnarray}
The cross-correlation spectrum of density $\delta$ with velocity
divergence $\theta= \nabla \cdot \mathbf{v}$ becomes
\footnote{While deriving this equation, we used the differential
form of continuity equation~(\ref{eq:vg0}), i.e. $ \partial
\delta/\partial t +a^{-1} \nabla \cdot \mathbf{v}=0$. We transform
this into Fourier space and notice that ``$\nabla \cdot
\mathbf{v}$'' in Fourier space is $-i\mathbf{k} \cdot \mathbf{v}$,
thus obtain Eq.~(\ref{eq:Ptheta-delta}).}
\begin{eqnarray}
P_{\theta \delta}&=&\langle (H f \delta_{\rm m}(k)) \delta_{\rm
g}(k) \rangle  \notag
\\
&=& b_{\mathrm{tot}}(k)Hf P_{\rm m}(k) .\label{eq:Ptheta-delta}
\end{eqnarray}

\begin{figure*}
\centerline{\includegraphics[bb=0 0 523
353,width=3.3in]{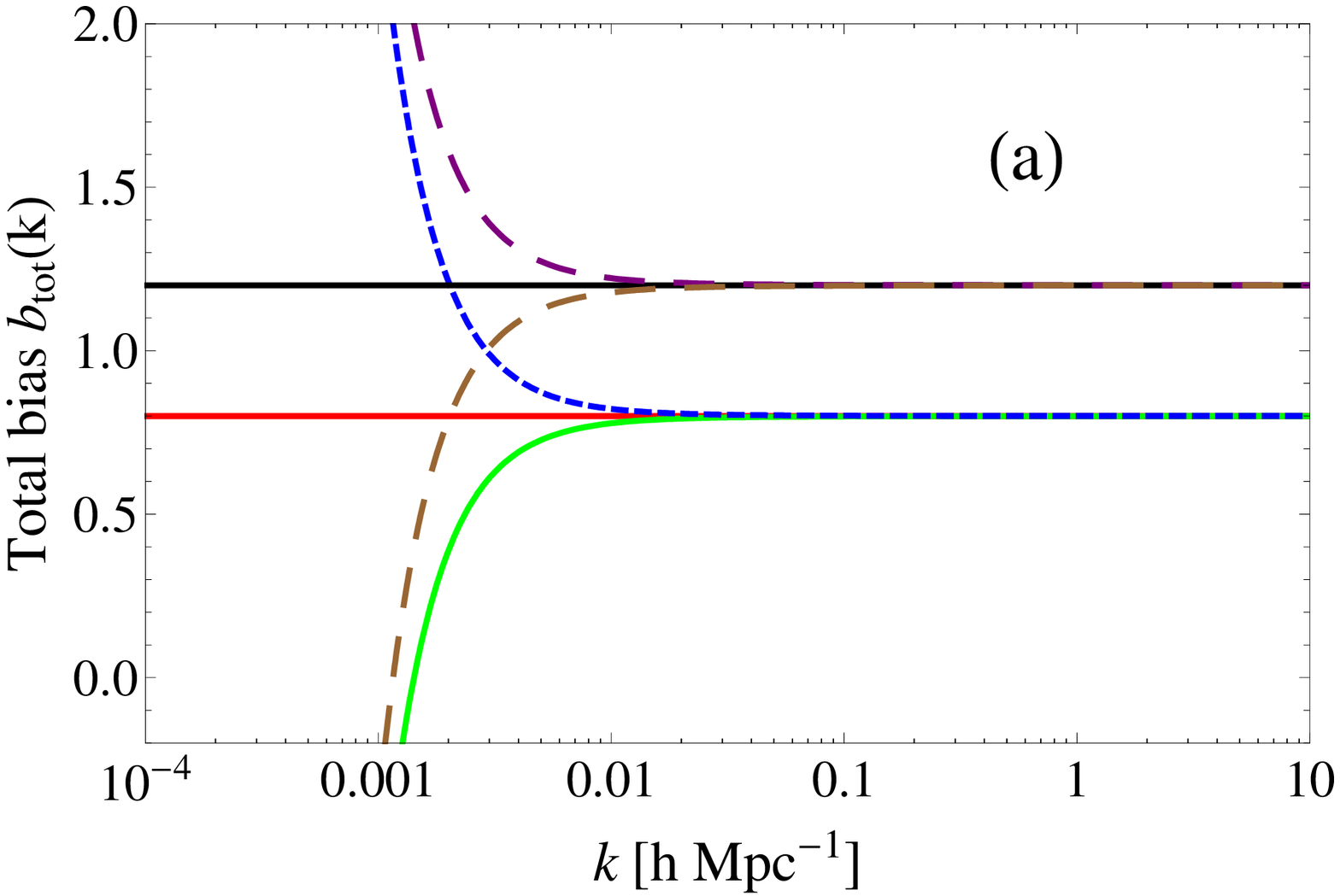}
\includegraphics[bb=0 0 592 385,width=3.4in]{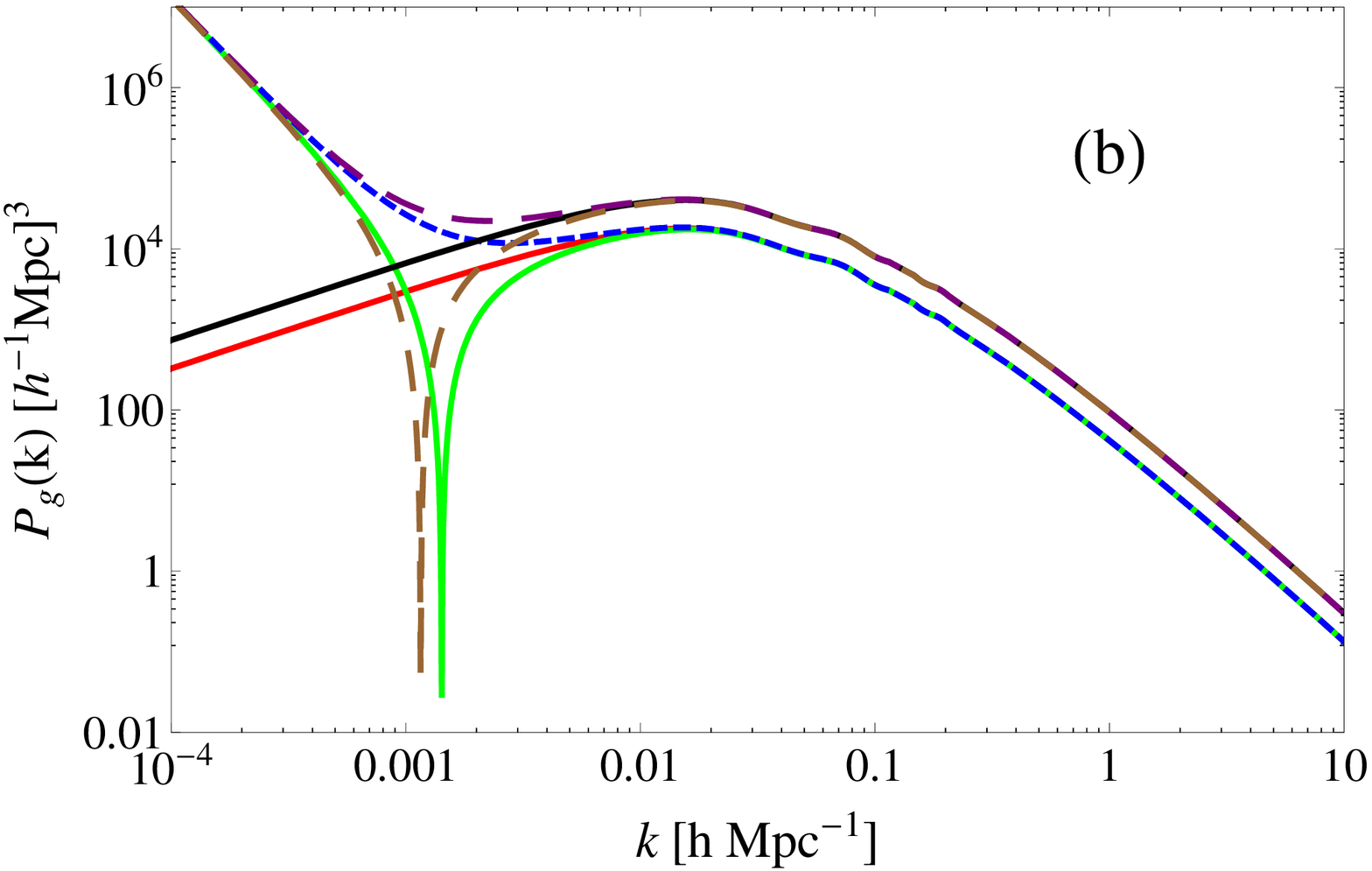}}
\centerline{\includegraphics[bb=0 0 761
384,width=4.8in]{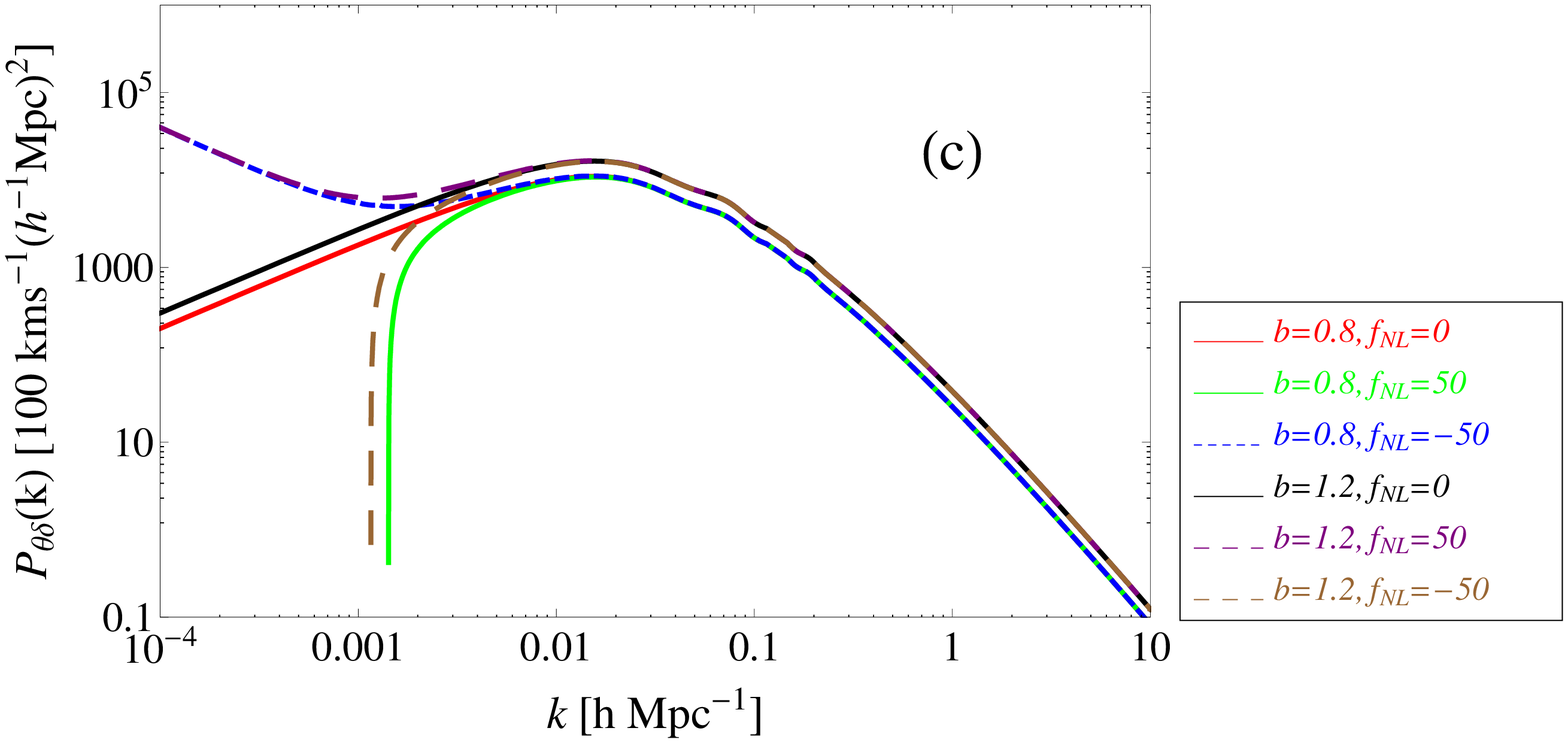}} \caption{\textit{(a)}: Total
bias (Eq.~(\protect\ref{eq:btot})) as a function of $k$.
\textit{(b)}: Galaxy power spectrum. \textit{(c)}: Galaxy and
velocity divergence cross-correlation power spectrum. The legend
for the colour scheme is shown on the right-hand side of panel
(c).} \label{fig:btot-pk}
\end{figure*}

In Figs.~\ref{fig:btot-pk}(a)--(c), we plot the total bias and
galaxy power spectrum as a function of $k$ over the range
$10^{-4}$ to $10 h\,\text{Mpc}^{-1}$.
In Fig.~\ref{fig:btot-pk}a,   comparing to the case of $f^{\mathrm{%
local}}_{\mathrm{NL}}=0$, one can see that either positive or
negative $\fnl$ tends to bias the galaxy power spectrum on very
large scales, but whether the bias is positive or negative depends
also on the value of $b$. Note that the non-Gaussian correction to
the bias [the function $A(k)$ in Fig.~\ref{fig:ak}] can completely
dominate the Gaussian bias at large scales. Therefore in
Fig.~\ref{fig:btot-pk}, we choose six sets of parameters to
represent different values of Eulerian bias $b$ and
non-Gaussianity $\fnl$ for different initial conditions of the
fluctuations. We describe each of these below.

\begin{enumerate}

\item $\fnl=0$ and $b>1$ (black solid lines). The primordial
perturbations are Gaussian, and $b>1$ so the objects are more
clustered than matter, and as a result they more easily form large-scale
overdensities.

\item $\fnl=0$ and $b<1$ (red solid lines). Here the primordial
perturbations are Gaussian, but the objects are anti-correlated
with respect to the matter fluctuations. The haloes with
anti-Lagrangian bias ($b_{\rm L}<0$) are always low mass objects,
well below the typical halo mass ($\sim 10^{10}\msun$). Such
low-mass haloes only survive today if they have avoided being
incorporated into more massive objects, and thus they
preferentially avoid high-density regions and therefore they are
anti-biased. Such an anti-bias can suppress the strength of the
galaxy power spectrum at all scales (red solid line in
Fig.~\ref{fig:btot-pk}b).

\item $\fnl>0$ and $b>1$ (purple long-dashed lines). In this
case, objects such as galaxies are more clustered than the matter distribution,
and small-scale fluctuations in the galaxy distribution are
positively correlated with large-scale fluctuations, so that $b_{\rm{tot}}$
at large scales becomes a large positive number. This enhances the
galaxy power spectrum ($P_{\rm gg}$) and galaxy-velocity
divergence ($P_{\theta \delta}$) power spectrum at large scales.

\item $\fnl>0$ and $b<1$ (green lines). When $b<1$ this means that
the primordial haloes and peaks are anti-biased with respect to
matter fluctuations. Non-Gaussianity with positive $\fnl$
generates a correlation between the small-scale modes that form
haloes and large-scale modes. The second term of
Eq.~(\ref{eq:btot}) therefore suppresses the clustering of haloes
on some certain scales. In addition, one can see that at $k \sim
0.0015 \mpch$, the total bias $b_{\rm{tot}}(k)$
(Eq.~(\ref{eq:btot})) vanishes, so that haloes are uncorrelated
with matter, producing a nearly-vanishing galaxy power spectrum at
this scale (the spike in Fig.~\ref{fig:btot-pk}b).

\item $\fnl<0$ and $b>1$ (brown long-dashed lines). This
corresponds to the case where the galaxies formed are related to
peaks in the primordial matter fluctuation, but the negative
$\fnl$ means that those small peaks are associated with
large-scale troughs of matter fluctuations. The total
bias is therefor suppressed at large scales.

\item $\fnl<0$ and $b<1$ (blue dashed lines). Here the galaxy
overdensities are anti-correlated with the matter fluctuations, so
the galaxies form in the troughs of the primordial density field.
However since $\fnl<0$, these troughs are associated with the
large-scale peaks of the galaxy distribution. Therefore, the fact
that galaxy small-scale overdensities and large-scale modes are
both anti-correlated with small-scale density fluctuations mean
that they are positively correlated with each other. Therefore one
obtains a positive total bias factor between $\delta_{\rm g}$ and
$\delta_{\rm m}$.

\end{enumerate}

No matter which subcases that the initial conditions fall into, on
the very large scales ($k \lesssim 10^{-4} \mpch$) the galaxy
power spectra with $\fnl \neq 0$ all converge with each other.
This indicates that at very large scales, the primordial
non-Gaussianity induced correlation always dominates the
clustering properties of galaxies.

\begin{figure*}
\centerline{\includegraphics[bb=0 0 600
390,width=3.3in]{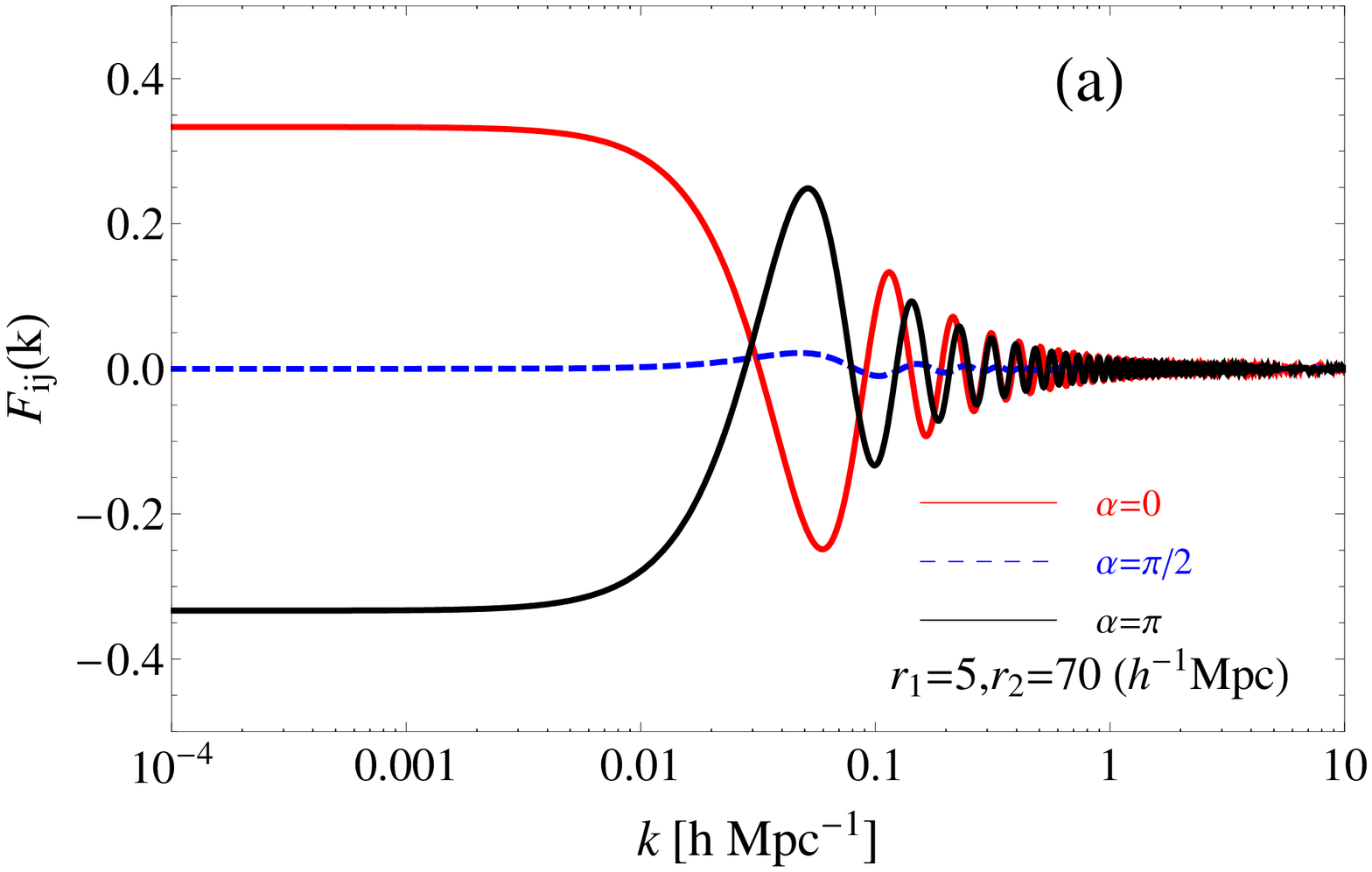}
\includegraphics[bb=0 0 656 423,width=3.3in]{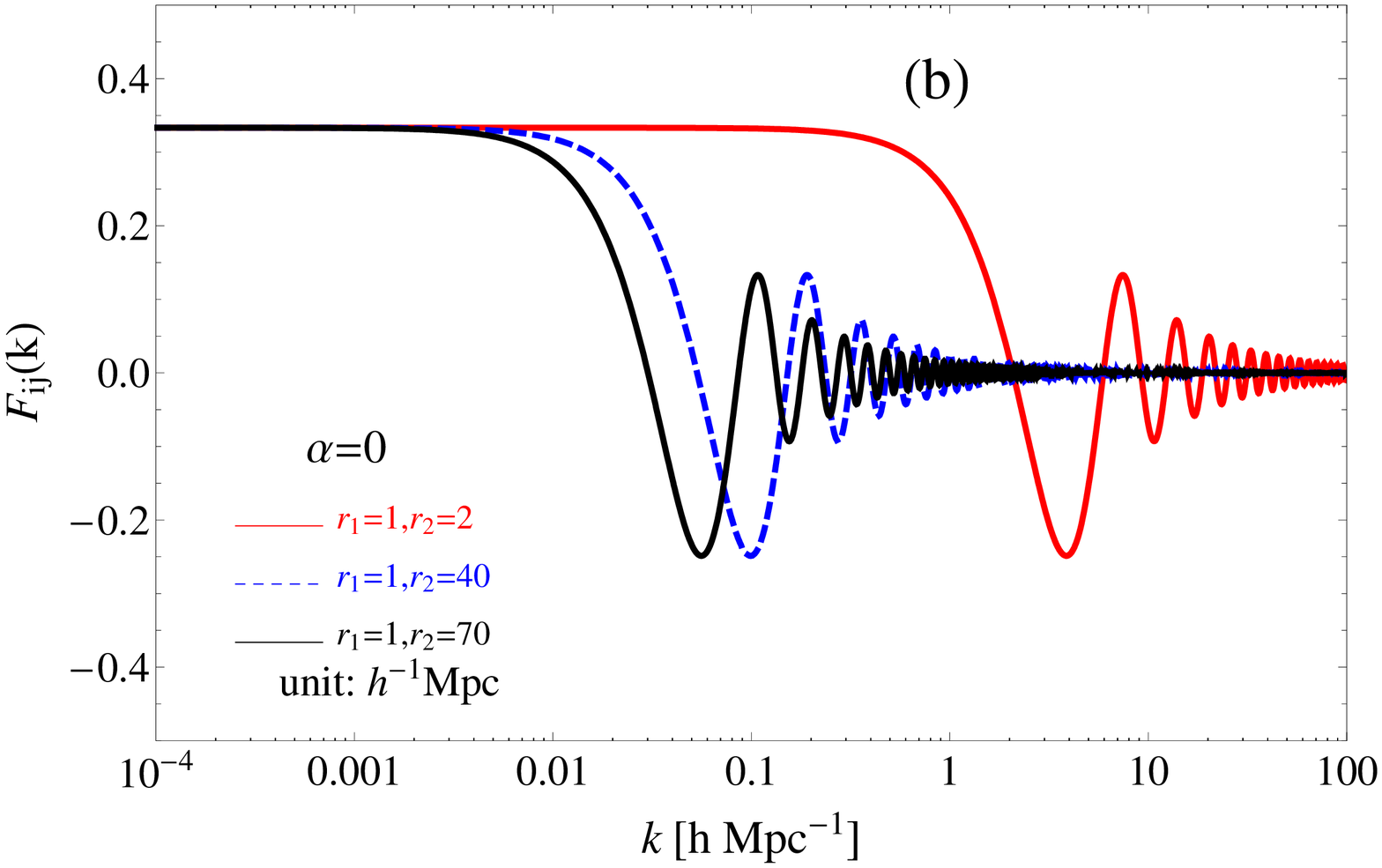}}
\caption{Angular dependence function $F_{ij}(k)$ for different
choices of the length of vectors $r_i = r_1, r_j = r_2$ and their
angular separation $\alpha$. In panel (a), we
fix the lengths of two vectors and plot $F_{ij}(k)$ with three
values of the separation angle $\alpha$, while in panel (b), we fix $\alpha=0$
and vary the length of the second vector $r_{2}$.} \label{fig:fij}
\end{figure*}

\subsection{Constraining $\fnl$ with peculiar velocities}

\label{sec:statistics} Now let us calculate the peculiar velocity
field and reconstructed density field in the framework of
non-Gaussian initial fluctuations. The real-space 3D velocity
field (Eq.~(\ref{eq:vg0})) can
also be expressed in Fourier space \footnote{%
In the following derivation, we use the bold character
$\mathbf{k}$ to express the 3D vector, and normal characters $k$
and $\hat{k}$ to express the norm and direction of $\mathbf{k}$,
i.e. $k=\sqrt{\mathbf{k\cdot k}}$, $\mathbf{k}=k\hat{k}$.} as
\begin{equation}
\mathbf{v}(\mathbf{r})=\frac{iH_{0}f}{(2\pi )^{3}}\int
{\rm d}^{3}\mathbf{k}\delta_{\rm m}(\mathbf{k})\frac{\mathbf{k}}{k^{2}}\exp \left( i\mathbf{k\cdot r}%
\right). \label{eq:vx}
\end{equation}
Now we substitute $\delta_{\rm
g}(\mathbf{k})=b_{\rm{tot}}(k)\delta_{\rm m}(\mathbf{k})$ into the
above equation, and after some arrangement, we have
\begin{eqnarray}
\mathbf{v}(\mathbf{r}) &=& \frac{iH_{0}f}{(2\pi )^{3}}\int {\rm
d}^{3}\mathbf{k} \frac{\delta
_{\rm g}(\mathbf{k})}{b_{\rm{tot}}(k)}\frac{\mathbf{k}}{k^{2}}\exp \left( i\mathbf{k\cdot r}%
\right) \nonumber \\
&=& \frac{iH_{0}f}{(2\pi)^{3}} \int {\rm d}^{3}\mathbf{k}
\left(\frac{1}{b}+\frac{1}{b_{\rm{tot}}(k)} -\frac{1}{b} \right)
\delta_{\rm g}(\mathbf{k}) \frac{\mathbf{k}}{k^{2}} \exp\left(i
\mathbf{k} \cdot \mathbf{r} \right) \nonumber \\
&=& \frac{iH_{0}f}{(2\pi)^{3}} \int {\rm d}^{3}\mathbf{k}
\frac{1}{b} \delta_{\rm g}(\mathbf{k}) \frac{\mathbf{k}}{k^{2}}
\exp\left(i
\mathbf{k} \cdot \mathbf{r} \right) \nonumber \\
&+& \frac{iH_{0}f}{(2\pi)^{3}} \int {\rm d}^{3}\mathbf{k} \left(
\frac{b - b_{\rm{tot}}(k)}{b \cdot b_{\rm{tot}}(k)}\right)
\delta_{\rm g}(\mathbf{k}) \frac{\mathbf{k}}{k^{2}} \exp\left(i
\mathbf{k} \cdot \mathbf{r} \right). \label{eq:vx2}
\end{eqnarray}
Absorbing the linear reconstruction $\beta=f/b$ into the first
term, performing a Fourier transform, and substituting
Eq.~(\ref{eq:btot}) into the second term, we obtain
\begin{eqnarray}
\mathbf{v}(\mathbf{r}) &=& \frac{H_{0}\beta}{4 \pi} \int {\rm
d}^{3}\mathbf{r'} \delta_{\rm g}(\mathbf{r'})
\frac{\mathbf{r'}-\mathbf{r}}{|\mathbf{r'}-\mathbf{r}|^{3}}
\nonumber \\
&-& \frac{iH_{0}(f/b)}{(2\pi)^{3}} \int {\rm d}^{3}\mathbf{k}
(\Delta b(k)) \delta_{\rm m}(\mathbf{k}) \frac{\mathbf{k}}{k^{2}}
\exp\left(i \mathbf{k} \cdot \mathbf{r} \right), \label{eq:vx3}
\end{eqnarray}
in which we can see that the primordial non-Gaussianity produces
an additional term in the velocity field.

Now we consider the meaning of the two terms in
Eq.~(\ref{eq:vx3}). The first term is the linear peculiar velocity
field reconstruction from observed galaxy distribution, which in
the following section, will be represented by the \textit{IRAS}
1.2\,Jy and PSC$z$ (Point Source Catalogue redshift) samples
\citep{Fisher1995,Saunders00}. The second part is the additional
3D velocity term coming from the non-Gaussian structures which
arise from the presence of local non-Gaussianity. This will appear
as a \textit{residual velocity field} $v^{\rm res} = v^{\rm mea} -
\beta v^{\rm rec}$ if we subtract the linearly reconstructed field
from the observed one. Since we can only measure the line-of-sight
velocity of distant galaxies, we need to project
Eq.~(\ref{eq:vx3}) on to the radial direction. We define the
projected left-hand-side of Eq.~(\ref{eq:vx3}) as
$v^{\rm{mea}}_{i}=\mathbf{v}(\mathbf{r})\cdot \hat{r}_{i}$ for the
$i\rm{th}$ object, since this represents the measured
line-of-sight velocity. We then define the projected first term in
the right-hand-side as $\beta v^{\rm{rec}}_{i}$ for the
reconstructed velocity for the $i\rm{th}$ object, where
$v^{\rm{rec}}_{i}$ is the reconstructed velocity with
normalization $\beta=1$. Now we can move the first term from the
right hand side to the left and calculate the covariance matrix
for the residual velocities $v^{\rm res}_i, v^{\rm res}_j$:
\begin{eqnarray}
C_{ij} &=&  \left \langle v^{\rm res}_i v^{\rm res}_j\right\rangle \nonumber\\
&=& \left \langle (v^{\rm{mea}}_{i}-\beta v^{\rm{rec}}_{i})
(v^{\rm{mea}}_{j}-\beta v^{\rm{rec}}_{j}) \right \rangle \nonumber
\\ &=& \frac{H_{0}^2(f/b)^2}{(2\pi)^{6}} \int {\rm d}^{3}\mathbf{k_{1}}
{\rm d}^{3}\mathbf{k_{2}}
 (\Delta b(k))^{2} \langle
\delta_{\rm m}(\mathbf{k}_{1}) \delta_{\rm m}(\mathbf{k}_{2})
\rangle \nonumber
\\
& \times & \frac{(\mathbf{k_{1}}\cdot
\hat{r}_{i})(\mathbf{k_{2}}\cdot \hat{r}_{j})}{(k_{1}k_{2})^{2}}
\exp \left(i (\mathbf{k_{1}}\cdot
\mathbf{r}_{i}-\mathbf{k_{2}}\cdot \mathbf{r}_{j}) \right) \nonumber \\
&&\ +\ \delta_{ij}(\sigma^{2}_{i}+\sigma^{2}_{\ast}) , \nonumber \\
\label{eq:variance1}
\end{eqnarray}
where $\sigma_{i}$ is the measurement error of the $i\rm{th}$
object, $\delta_{ij}$ is the kronecker delta symbol, and
$\sigma_{\ast}$ is the intrinsic small-scale velocity dispersion.
Substituting the ensemble average of the matter density contrast
$\langle \delta_{\rm m}(\mathbf{k}_{1}) \delta_{\rm
m}(\mathbf{k}_{2}) \rangle
=(2\pi)^{3}\delta^{3}(\mathbf{k}_{1}-\mathbf{k}_{2})P_{\rm{m}}(k)$,
and Eq.~(\ref{eq:Bfnl}), we obtain
\begin{eqnarray}
C_{ij} =\frac{H^{2}_{0}}{2\pi^{2}} \left(f-\beta \right)^{2}
I_{ij}+ \delta_{ij}(\sigma^{2}_{i}+\sigma^{2}_{\ast}),
\label{eq:variance2}
\end{eqnarray}
where
\begin{eqnarray}
I_{ij}= \left(f^{\rm{local}}_{\rm{NL}}\right)^2 \int {\rm d}k
P_{\rm{m}}(k)A^{2}(k)F_{ij}(k), \label{eq:Iij}
\end{eqnarray}
and $F_{ij}(k)$ is the integral of angle over the full-sky
\begin{eqnarray}
F_{ij}(k)=\frac{1}{4\pi} \int {\rm d} \Omega_{k}(\hat{k}\cdot
\hat{r}_{i}) (\hat{k}\cdot \hat{r}_{j}) \exp
\left(i\mathbf{k}\cdot \left(\mathbf{r}_{i}-\mathbf{r}_{j} \right)
\right), \label{eq:Fij}
\end{eqnarray}
which can be calculated analytically. In the appendix of
\cite{Ma11} it is shown that
\begin{eqnarray}
F_{ij}(k) = \frac{1}{3}\cos \alpha
(j_{0}(k\tilde{r})-2j_{2}(k\tilde{r})) +  \frac{%
1}{\tilde{r}^{2}}j_{2}(k\tilde{r})r_{i}r_{j}\sin ^{2}\alpha ,
\label{eq:fij2}
\end{eqnarray}
with
\begin{eqnarray}
\tilde{r}=\left|\mathbf{r}_{i}-\mathbf{r}_{j}\right|=\left[r^{2}_{i}+r^{2}_{j}-2r_{i}r_{j}\cos(\alpha)
\right]^{1/2}. \label{eq:funcA}
\end{eqnarray}
From Eqs.~(\ref{eq:fij2}) and (\ref{eq:funcA}), one can see that
$F_{ij}(k)$ only depends on three values: the length of vectors
$\mathbf{r}_{i}$ and $\mathbf{r}_{j}$ (i.e. $r_{i}$ and $r_{j}$),
and their separation angle $\alpha$ ($\cos(\alpha)=\hat{r}_{i}
\cdot \hat{r}_{j}$). In Fig.~\ref{fig:fij}, we plot the function
$F_{ij}(k)$ by taking different sets of parameter values. In
Fig.~\ref{fig:fij}a, we fix the length of the two vectors $r_{1}=5
\hmpc$ and $r_{2}=70 \hmpc$, and vary the separation angle
$\alpha$. One can see that different $\alpha$ values determine the
amplitude of the function $F_{ij}$. As $\alpha$ increases from $0$
to $\pi$, the amplitude of $F_{ij}$ in the low-$k$ plateau flips
from positive to negative, and its value is close to zero if
$\alpha \rightarrow \pi/2$. However, no matter what $\alpha$ value
is taken, $F_{ij}(k)$ oscillates very quickly and converges to
zero at $k>0.02 \mpch$. In Fig.~\ref{fig:fij}b, we fix $\alpha=0$
and plot $F_{ij}(k)$ by varying the length of the second vector
$r_{2}$. One can see that if $r_{1}$ is close to $r_{2}$, the
function starts to dominate at larger $k$ (basically $k>1 \mpch$);
on the other hand, if $r_{1}$ and $r_{2}$ differ by a large value,
then the function will only be constant for $k<0.02\mpch$ and
will oscillate quickly and converge to zero at large $k$ values.

The shape of the function $F_{ij}(k)$ is important because the
covariance between velocities of different objects on the sky
(Eqs.~(\ref{eq:variance2}) and (\ref{eq:Iij})) is the matter power
spectrum filtered by the functions $F_{ij}(k)$ and $A^{2}(k)$.
$F_{ij}(k)$ oscillates and converges to zero very quickly at large
$k$, and $A(k)$ also decays very rapidly when the value of $k$
becomes larger (Fig.~\ref{fig:ak}). Therefore, the total
covariance $I_{ij}$ is sensitive to the large-scale behaviour of
the filtered function $P_{\rm{m}}(k)$, i.e. $I_{ij}$ is sensitive
to scales $k<0.02 \mpch$ and does not depend on the small-scale
modes of the matter power spectrum very much. In this sense, the
covariance matrix method proposed above is valid for the detection
of non-Gaussianity because as we see in Fig.~\ref{fig:btot-pk},
the major signature of primordial non-Gaussianity is in the very
large-scale galaxy density power spectrum.

Although most of the effect of non-Gaussianity is on large scales,
our filter function retains some sensitivity to smaller scales.
Fig.~\ref{fig:fij}a, for instance, shows that the correlation
between two galaxies is larger for small values of the separation
angle $\alpha$. In Fig.~\ref{fig:fij}b, we see that for small
separation angle and small difference in the radial distance ($r_1
\sim r_2$), the correlation signal can extend down to
$k=1$--$2\mpch$, although non-linear corrections and errors in
distance and velocity will become important on these small scales.
In comparison, the CMB probes non-Gaussianity to $\ell = 1500$,
which corresponds to structures with scales $0.0264$ Mpc at the
last scattering surface, evolving to $\sim 30$ Mpc at the present
day. This CMB sensitive scale corresponds to $k \sim 0.03 \mpch$,
while we can go down to $1 \mpch$, so we can extend these
constraints by an order of magnitude in $k$, albeit with poor
sensitivity.

\begin{figure}
\centerline{\includegraphics[bb=0 0 625
385,width=3.2in]{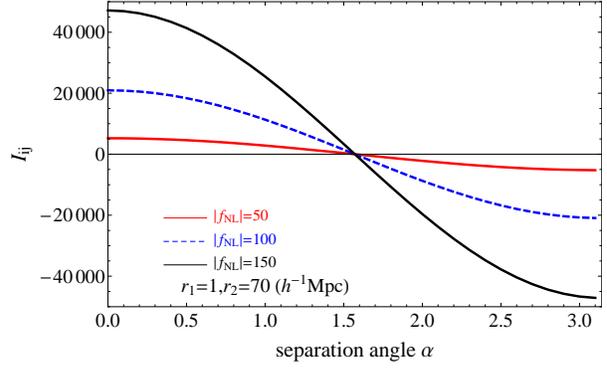}} \caption{Velocity condition quantity
$I_{ij}$ (Eq.~(\ref{eq:Iij})) as a function of separation angle
$\alpha$ when choosing different $\fnl$ parameters. One can see
that if the separation angle between two objects is less (greater)
than $\pi/2$, the correlation is positive (negative), and the
amplitude of the correlation is proportional to the magnitude of
the local non-Gaussianity.} \label{fig:Iij}
\end{figure}

We also plot (in Fig.~\ref{fig:Iij}) the quantity $I_{ij}$ as a
function of separation angle $\alpha$, while varying the $\fnl$
parameter. One can see that if $\alpha<\pi/2$, the two separated
objects tend to be correlated, while if $\alpha>\pi/2$, they tend
to be anti-correlated. The amplitude of the correlation is
proportional to the magnitude of local non-Gaussianity $\fnl$. In
addition, the shape of the correlation function $I_{ij}$, as a
function of $\alpha$, is very similar to a cosine function. This
important feature will allow us to obtain robust estimates of the
$\fnl$ value, for the following reason. In the case of ordinary
spiral galaxies, the dispersion $\sigma_{\ast}$ accounts for the
small-scale non-linear motions which are believed to have a
variance around $300 \kms$ \citep{Wang07}. We will assume this
value in the following likelihood analysis procedure. However, we
can see that the assumption of this particular value does not make
much difference in estimating the absolute magnitude of $\fnl$.
This is because, as seen in Fig.~\ref{fig:Iij}, $I_{ij}$ is only
sensitive to the angular separation of the two objects in
different directions, so this modulation function is close to a
first order polynomial function $P_{1}(\alpha)=\cos(\alpha)$,
which is orthogonal to the ``monopole'' moments of the correlation
($\sigma_{\ast}$). Therefore, it is the \textit{spatial
correlation} between different directions of the objects that
really constrains $\fnl$.

Therefore returning to Eq.~(\ref{eq:variance2}), one can see that
if we subtract the density-reconstructed $v^{\rm{rec}}_{i}$ from
the observed peculiar velocity field, covariance in the residual
field will consist of three parts: the primordial non-Gaussianity
induced cross-correlation between different velocities, which is
proportional to $I_{ij}$; the measurement error of the
line-of-sight velocity $\sigma_{ij}$; and the intrinsic
small-scale velocity dispersion $\sigma_*$. If there is no
non-Gaussianity, i.e. $\fnl=0$, then $I_{ij}=0$ and the first term
of the right-hand-side of Eq.~(\ref{eq:variance2}) vanishes, so
there is no correlation between different directions of the
residual velocity field. This corresponds to the Gaussian case,
where the residual field is absolutely randomly distributed across
the whole sky. The covariance matrix is then diagonal and is
determined only by the measurement errors and the small-scale
velocity and intrinsic dispersion terms\footnote{The
$\sigma_{\ast}$ value can include the unaccounted systematics of
the measurement error $\sigma_{i}$, similar to the
``hyper-parameter'' method used in \cite{Ma12a}.}.

On the other hand, if the \textit{residual map} shows
correlations in velocity from one region of the sky to another, it
may indicate the presence of primordial non-Gaussianity at some
level. Thus the covariance of the residual map will provide a quantitative
measure of the effects of local non-Gaussianity.
To constrain $f_{NL}^{\rm local}$, we can formulate a
likelihood function
\begin{eqnarray}
&& L(\beta,\fnl) =  \frac{1}{\sqrt{\det(C)}} \times \nonumber \\
&&
 \exp \bigg\{ -\frac{1}{2} \left(v^{\rm{mea}}_{i}-\beta
v^{\rm{rec}}_{i} \right) C_{ij}^{-1} \left(v^{\rm{mea}}_{j}-\beta
v^{\rm{rec}}_{j} \right) \bigg\},\label{eq:like1}
\end{eqnarray}
where $\fnl$ is contained in $C$ and $\beta$ is contained in both
$C$ and the data vector. In the following, we will apply this
likelihood function to a peculiar velocity field data set to
constrain the values of $\beta$ and $\fnl$.

\section{Data}
\label{sec:data} From the derivation above we see that in
order to quantify the primordial non-Guassianity present in the
primordial density field, we need to subtract from the observed peculiar
velocity field a model reconstructed from the density field assuming
linear bias, and measure the variance of the residual
velocities in different directions. In this section, we first
introduce the observed peculiar velocity field data sets and then
the model velocity field data set.

\subsection{The observed peculiar velocity field}
\label{sec:peculiar}

\begin{figure*}
\centerline{\includegraphics[bb=-163 76 775
715,width=3.4in]{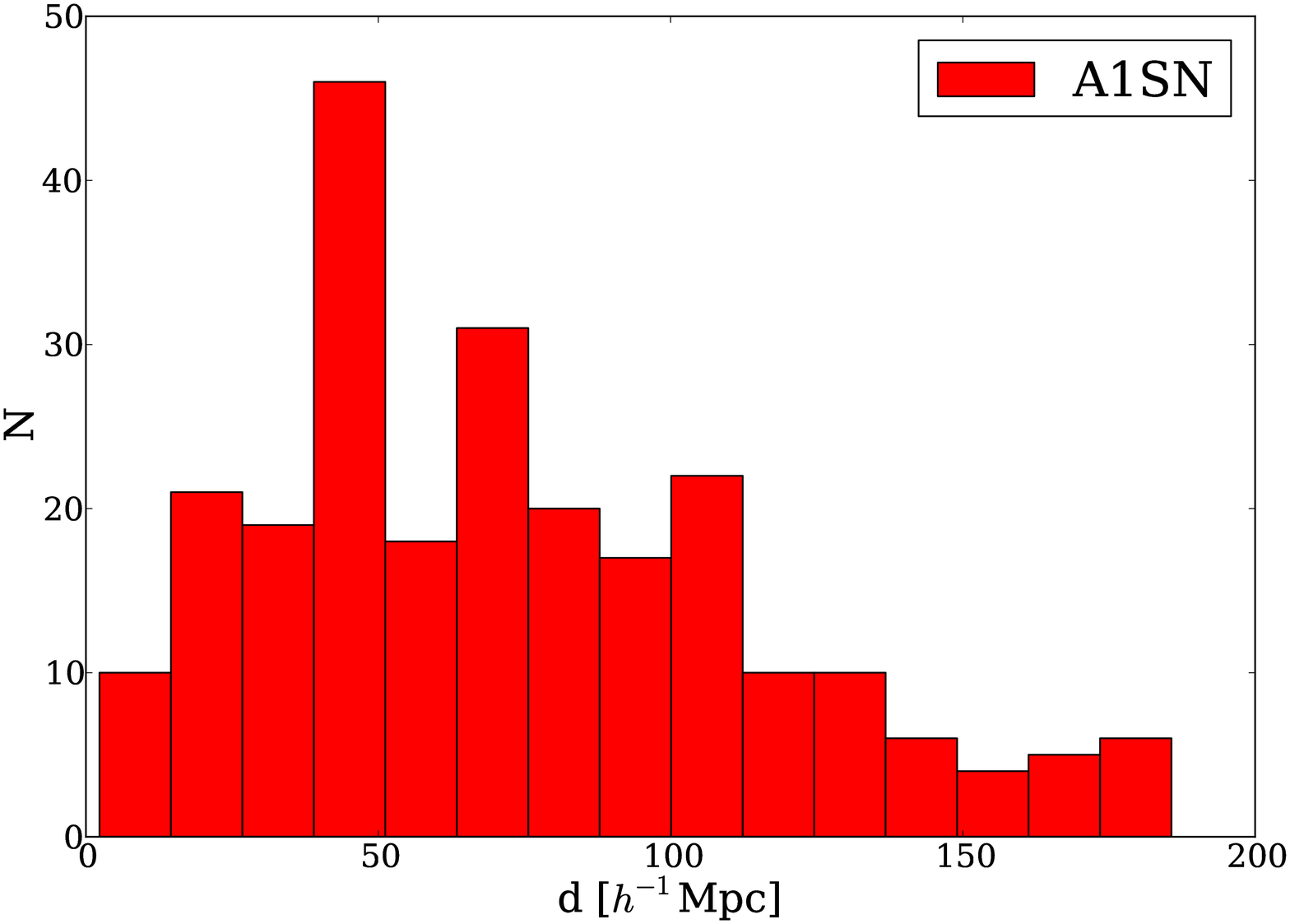}
\includegraphics[bb=-150 102 762 689,width=3.6in]{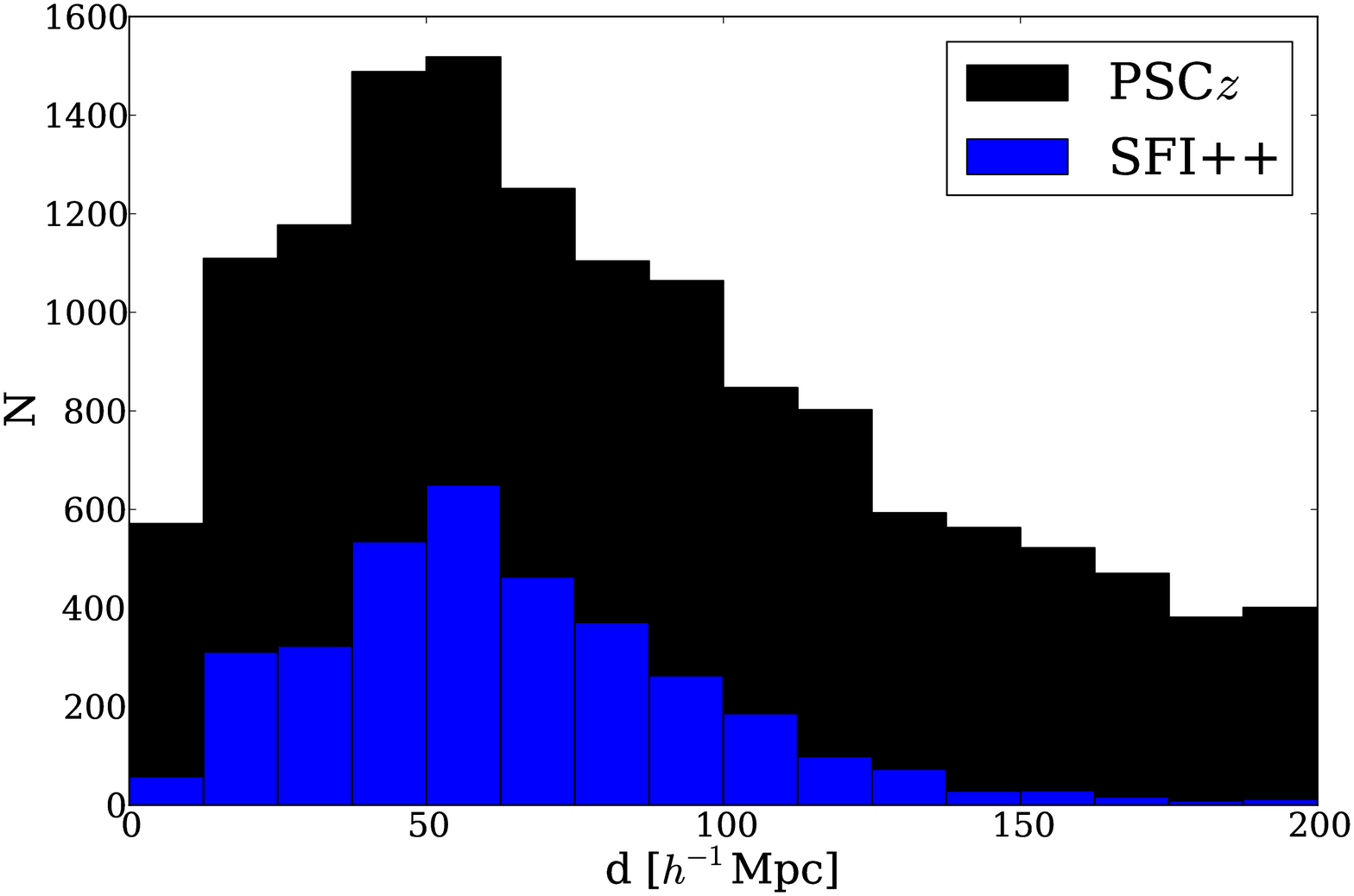}}
\caption{Distance distributions for the A1SN (left-hand panel),
SFI++ and PSC$z$ (right-hand panel) catalogues. }
\label{fig:histogram}
\end{figure*}

The observed peculiar velocity field data set is of course the
most important ingredient of the data analysis. In this work we
focus on two different catalogues, A1SN and SFI++. There are
several reasons for choosing these two catalogues. First, they are
high-quality and recently assembled peculiar velocity data sets
that deeply sample our local volume. Secondly, they both have
full-sky coverage, which allows us to constrain the correlation of
objects in different directions. Thirdly, the distance estimators
for the two data sets are completely independent of each other,
therefore minimizing the chance of systematic error, and providing
us a self-consistent check of the results.

The A1SN catalogue is known as the ``First Amendment'' supernovae
sample, and consists of $245$ Type-Ia supernovae compiled by
\cite{Turnbull12}. The data set is merged from three different
Type-Ia supernovae data sets: (1) $106$ samples from \cite{Jha07}
and \cite{Hicken09}; (2) another $113$ objects collected by
\cite{Hicken09}; and (3) the $28$ objects observed by the
``Carnegie Supernovae project'' \citep{Folatelli10}. Supernovae
observations use the luminosity-distance relation as the
``standard candle'', so the distance error is typically $7$ per
cent, much smaller than for the other galaxy Fundamental Plane or
Tully--Fisher-relation-determined distances. The characteristic
depth\footnote{This depth is defined as the inverse-error weighted
depth.} of the whole catalogue is $58 \hmpc$.

The Spiral Field \textit{I}-band (SFI++) catalogue is the largest
and densest survey of peculiar velocities available to date
\citep{Springob07}, and consists of $3456$ spiral galaxies with
peculiar velocities derived from the Tully--Fisher relation
\citep{Tully77}. Most of the galaxies are in the field (2675) or
in groups (726) \citep{Watkins09}. The distribution of the
galaxies across the sky is remarkably homogeneous, as shown in
fig.~1 of \cite{Feldman10} and \cite{Ma12a}. Since distances in
the catalogue are derived by using the Tully--Fisher relation, the
typical distance errors are around $23$ per cent. The
characteristic depth of the SFI++ catalogue is around
$40$-$50\hmpc$, as is shown in the right-hand panel of
Fig.~\ref{fig:histogram}.

In the left-hand panel of Fig.~\ref{fig:histogram}, we show the
distribution of distances in the A1SN sample, while in the
right-hand panel
 we plot the distributions for the PSC$z$ and SFI++ samples. By
comparing these three samples, we can see that the distribution of
A1SN and SFI++ is very close to a scaled down version of the
PSC$z$ sample, so that the shape of the distance distribution is
quite self-similar in each case. The three data sets also have
similar depths. This is another reason we choose the
A1SN and SFI++ data sets to compare with the PSC$z$ model velocities.

%
%
%
%

In general, these samples become sparser and errors increase at large distances.
 If we include the ``outliers'' in the range of $70$-$130\hmpc$, for instance,
 the scatter among $\beta$ values increases, indicating there may be
systematic errors in either the measured or model velocity field
\citep{Ma12a}. To be conservative, we exclude galaxies outside $70\hmpc$
and just use the samples within this range. With this cut, we retain
$126$ A1SN samples, and $2044$ SFI++ samples for the final data set.

As a last step, since we perform  the likelihood analysis in the
Local Group frame, we transform the velocities provided in the
CMB frame by subtracting the line-of-sight component of the Local
Group velocity determined from the CMB dipole, i.e. $v=611 \kms$
towards $(l,b)=(269^{\circ},+28^{\circ})$ (see \citealt{Scott10}).

\subsection{Model density and velocity fields}
\label{sec:model-vel}

In order to search for the effects of non-Gaussianity, we also
need a model velocity field. In this paper we use the model
velocity field obtained \citep{Branchini99} from the \textit{IRAS}
PSC$z$ catalogue \citep{Saunders00}. The spatial distribution of
PSC$z$ galaxies is fairly homogeneous across the sky
[cf.~\citealt{Ma12a}, fig.~1, panels (a) and (b)], which is ideal
for studying the cross-correlation with galaxies in different
directions.

PSC$z$ redshift catalogues were used to trace the underlying mass
distribution within $300\hmpc$, with the assumption of linear and
deterministic bias \citep{Radburn-Smith04,Ma12a}. The
reconstructed velocity field was obtained by performing the
integration of the first term of right hand side of
Eq.~(\ref{eq:vx3}) by using the iterative technique of
\cite{Yahil91}. The iteration procedure includes only objects
within $130\hmpc$, since at larger distances the samples
are very sparse
and do not strongly affect peculiar velocities within the $70
\hmpc$ volume. In this way, we obtain the final reconstructed
peculiar velocities for $8995$ PSC$z$ galaxies within $130 \hmpc$
that were not collapsed into galaxy clusters \citep{Ma12a}.

The distribution of distances in the PSC$z$ catalogue is plotted
in the right-hand panel of Fig.~\ref{fig:histogram}. Note that we
only use the $d<70\hmpc$ galaxies for the likelihood analysis, but
use the samples out to $130\hmpc$ to model the gravitational pull
of the distant structures.

Since the PSC$z$ predicted velocities outnumber the observed
peculiar velocities, we need to interpolate the predicted velocities
at the position of the galaxies in the peculiar velocity data sets
\citep{Ma12a}. We perform such ``smoothing'' by applying a
Gaussian kernel of the same radius $R_{j}\simeq 5 \hmpc$ to the
predicted velocity field, i.e. we calculate
\begin{equation}
\mathbf{v}_{\rm{smo}}(\mathbf{r}_{i})=\frac{\sum_{j=1}^{N}\mathbf{v}_{\rm{rec}}(\mathbf{r%
}_{j})\exp \left( -\frac{(\mathbf{r}_{j}-\mathbf{r}_{i})^{2}}{2R_{j}^{2}}\right) }{%
\sum_{j=1}^{N}\exp \left( -\frac{(\mathbf{r}_{j}-\mathbf{r}_{i})^{2}}{%
2R_{j}^{2}}\right) },
\end{equation}
where we sum over $N$ PSC$z$ galaxies at position $\mathbf{r}_{j}$
to interpolate to the position of galaxy ($\mathbf{r}_{i}$) in the
peculiar velocity catalogue. We then project the smoothed
velocities on to the line-of-sight direction to compare with the
observed velocity.

As we pointed out in our previous work \citep{Ma12a}, the typical
random errors of the model velocity field are $130 \kms$
\citep{Branchini99}, much smaller than the observed velocities. In
our earlier $v$--$v$ comparison, we initially gnored this random
error, but eventually found a relatively large value of a
``hyper-parameter'' which indicates the amplitude of
underestimated errors. In this work, we group all of these random
errors and unaccounted systematics into one parameter,
$\sigma_{\ast}$ (see Eqs.~(\ref{eq:variance1}) and
(\ref{eq:variance2})). Since the random errors of the
reconstructed velocities are around $130 \kms$, the thermal
velocities are typically $\sim 250 \kms$ \citep{Wang07}, and there
may be other systematics unaccounted for, we set the parameter
$\sigma_{\ast}=300 \kms$. Changing this parameter to be larger or
smaller value does not strongly affect the constraints on $\fnl$,
because these come mainly from the ``dipole'' modulation of the
covariance matrix while the $\sigma_{\ast}$ is a monopole term.

\section{Results}
\label{sec:results}


\begin{figure*}
\centerline{\includegraphics[bb=0 0 561
375,width=3.2in]{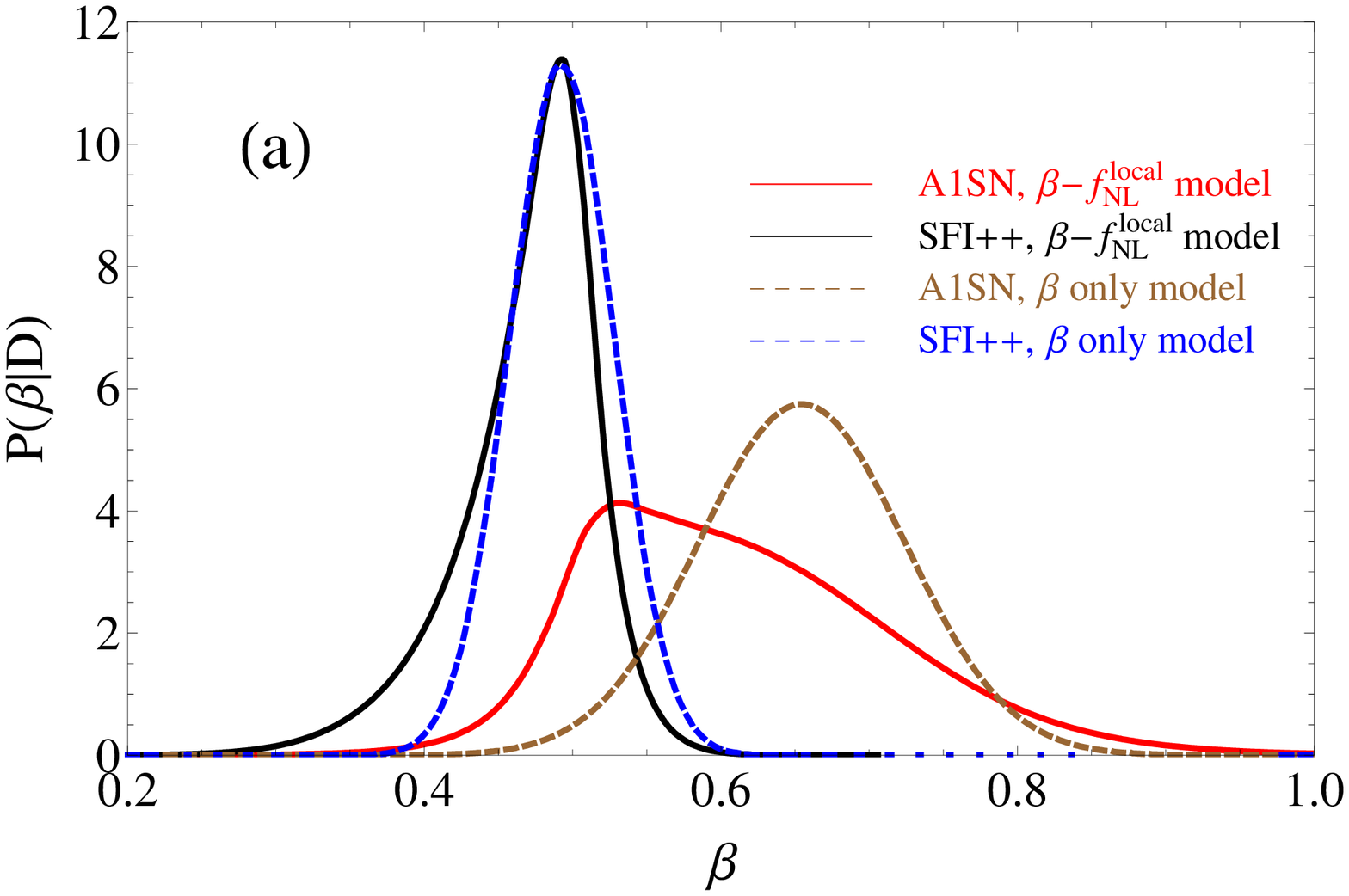}
\includegraphics[bb=0 0 548 358,width=3.3in]{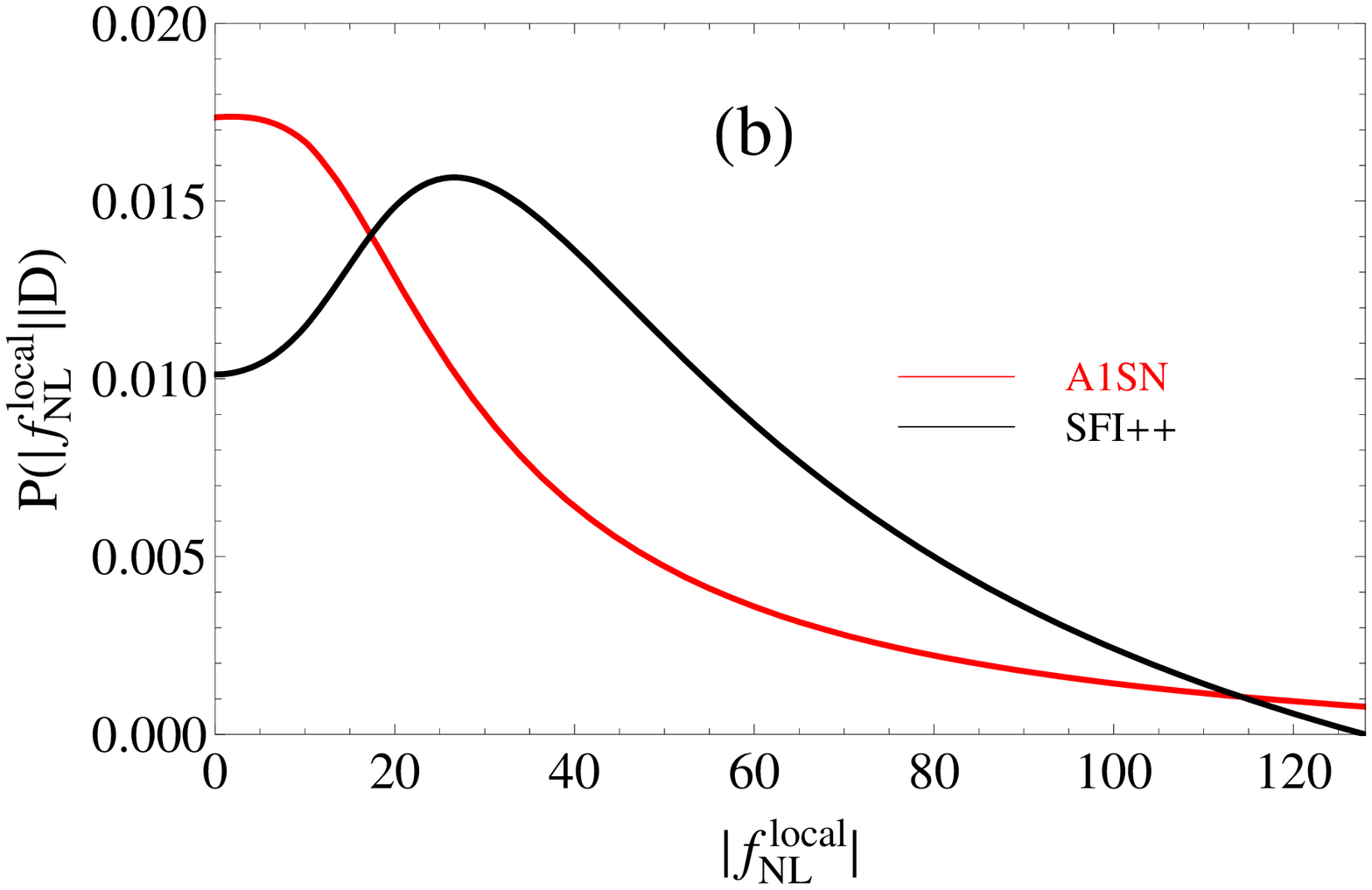}}
\caption{\textit{Left} -- marginalized distribution of the $\beta$
parameter with and without the correlation term in the covariance
matrix (Eqs.~(\ref{eq:variance1}) and (\ref{eq:variance2})).
\textit{Right} -- marginalized distribution of the $\fnlabs$
parameter for the model including non-Gaussianity-induced spatial
correlations.} \label{fig:like}
\end{figure*}

\begin{table*}
\renewcommand{\arraystretch}{1.5}
\begin{centering}
\begin{tabular}{|c|c|c|c|c|}
\hline Data set & Model & $\beta$ value & $\fnlabs$ value & $-\log
L_{\rm{min}}$ \\ \hline A1SN & $\beta$-$\fnl$ &
$0.53^{+0.15}_{-0.04}$ & $0.0 \pm 25.7$  & $681.7$ \\
\cline{2-5} & $\beta$-only & $0.65^{+0.07}_{-0.06}$ &  & $681.7$
\\ \hline
SFI++ & $\beta$-$\fnl$ & $0.49^{+0.03}_{-0.05}$ & $26.6\pm 33.0$ &
$14159.1$ \\ \cline{2-5} & $\beta$-only & $0.49^{+0.04}_{-0.03}$ &
 & $14159.1$
\\\hline
\end{tabular}%
\caption{The results of constraints from A1SN and SFI++ data sets.
All of the errors quoted are $1\sigma$ CL.} \label{tab1}
\end{centering}
\end{table*}


By substituting the observed line-of-sight velocities
$v^{\rm{mea}}_{i}$ and reconstructed velocities $v^{\rm{rec}}_{i}$
($i=1,...N$) into Eq.~(\ref{eq:like1}), we can obtain the joint
likelihood of $\beta$ and $\fnlabs$. Note that in
Eq.~(\ref{eq:Iij}), the covariance matrix is proportional to
$\fnlabs$, so the likelihood can only determine the absolute
magnitude of $\fnl$. In Fig.~\ref{fig:like}, we plot the
probability distribution of $\beta$ (panel a) and $\fnlabs$ (panel
b)  by using the A1SN and SFI++ catalogues. We also list our
results in Table~\ref{tab1}.

In Fig.~\ref{fig:like}a, we plot the probability distribution of
$\beta$. In this figure, we combine the results both from the
model which includes correlations between different directions in
the residual velocities (Eqs.~(\ref{eq:variance1}) and
(\ref{eq:variance2})) and the model just with measurement errors
and small-scale and intrinsic dispersion [i.e. only with second
term in Eqs.~(\ref{eq:variance1}) and (\ref{eq:variance2})]. In
Fig.~\ref{fig:like} and the following, we refer to these as the
$\beta$-$\fnl$ and the $\beta$-only models, respectively. One can
see that for the SFI++ data set, the best-fitting value of $\beta$
is $0.49$, close to the value we obtained from the
``hyper-parameter'' method \citep{Ma12a}. By neglecting the
correlation term, the peak of the distribution does not change,
but the width of the distribution is reduced. In this sense, the
additive covariance term from primordial non-Gaussianity
(Eqs.~(\ref{eq:variance1}) and (\ref{eq:variance2})) broaden the
distribution of the $\beta$ parameter, without shifting the
best-fitting value very much.

The same is true of the A1SN data set, except that the peak of the
distribution shifts a little towards a slightly lower $\beta$
value. For both the data sets, the $\beta$ values found are
consistent with each other, and are also consistent with the
values found in \cite{Ma12a}, indicating that the growth of
structure rate ($f\sigma_{8}$) is consistent with the prediction
of the $\Lambda$-Cold-Dark-Matter ($\Lambda$CDM) model.

In Fig.~\ref{fig:like}b, we plot the marginalized distribution of
$\fnl$, using the likelihood function (Eq.~(\ref{eq:like1})). One
can see that both data sets prefer the $\fnlabs$ value to be very
small, consistent with zero within $1\sigma$ CL. A1SN data prefers
the $\fnlabs<27$ at $68\%$ CL, and SFI++ data suggest
$\fnlabs<59.6$ at $68\%$ CL. These two upper bounds are the
strongest upper limits we can obtain from current measurements of
the local peculiar velocity field. Overall, our constraints on
$\fnl$ are tighter than any derived previously from large-scale
structure measurements (e.g.~\citealt{Xia11}, $\fnl=48\pm 20$ at
$1\sigma$ CL.; \citealt{Nik13}, $\fnl=90 \pm 30$ at $1\sigma$ CL.;
and \citealt{Ross13}, $\fnl<195$ at $1\sigma$ CL.). This
constraint is still considerably weaker than the recent result
from \cite{Planck24} ($\fnl=2.7\pm 5.8$), but it does cover a
slightly different range of scales, and in general provides an
independent confirmation at low redshift. The peculiar velocity
constraint could also be improved with more data from a future
velocity survey.

If there really was a primordial non-Gaussian correlation between
objects in different directions of the sky, the negative
log-likelihood function of the full covariance matrix would be
much lower than the diagonal one alone. This is not the case,
however. In Table~\ref{tab1}, we list the minimum value of $-\log
L$ for both of the likelihoods, with and without spatial
correlations. One can see that the two likelihood methods give the
same value of $-\log L$, suggesting that the two models provide
the same goodness of fit. However, since the full-covariance
matrix has one more parameter than the $\beta$-only model, the
current observational data from the peculiar velocity field and
model velocity field do not support strong evidence of large
$\fnl$.

\section{Conclusion}
\label{sec:conclude}

Primordial cosmological perturbations are usually assumed to have
Gaussian statistics, as expected in single-field inflation models.
Many variants on single-field inflation predict deviations from
Gaussianity, however, and there have been some tentative claims of
non-Gaussianity from previous observations of large-scale
structure \citep{Xia11,Nik13,Ross13}. Here we have shown that
measurements of local peculiar velocities and density fields set
strong constraints on departures from Gaussian initial conditions
down to scales 10 times smaller than those probed by the CMB.

The peculiar velocity field of galaxies traces the underlying
matter distribution directly, whereas the galaxy density
distribution may have a scale-dependent bias with respect to the
matter distribution if the initial conditions are (locally)
non-Gaussian. The peculiar velocity field can be decomposed into
two terms, the velocity field reconstructed assuming linear bias
(i.e.~the usual Gaussian term) and, a second, \textit{residual
velocity field}. For Gaussian initial conditions, the residual
velocity field should be random, without any large-scale spatial
correlations. If $\fnl$ is non-zero, however, scale-dependent bias
in the galaxy distribution will induce large-scale spatial
correlations in the residual velocity field. We construct a
likelihood function to quantify the variance of angular
correlations in the residual map, and thereby constrain $\fnl$.

Applying our likelihood function to the currently available deep,
full-sky surveys of Type-Ia supernovae (A1SN) and spiral galaxies
with Tully--Fisher determined distances (SFI++), we find that
models with and without local non-Gaussianity give consistent
constraints on $\beta$, constraints which are also consistent with
our previous work \citep{Ma12a}. This also confirms that the
linear growth rate at the present time is consistent with the
predictions for the $\Lambda$CDM model with \textit{WMAP} and
\textit{Planck} determined cosmological parameters.

More importantly, we can also constrain the amplitude of local
non-Gaussianity by comparing the log-likelihood for the models
with and without an $\fnl$ term. This analysis provides an upper bound of
$\fnlabs \leq 25.7$ (A1SN), and $\fnlabs \leq 59.6 $ (SFI++) at $1\sigma$ CL.
These limits are as tight as any set by previous large-scale structure studies.
We find that models with or without non-Gaussianity provide the same
``goodness'' of fit, indicating that adding the spatial correlation parameter
does not improve the fit to the residual velocity field.

Although we do not find any signature of primordial
non-Gaussianity, our physical and statistical model is an
independent constraint on the primordial initial conditions,
complementary to the CMB measurements. This method is applicable
to any data set with peculiar velocities and an overlapping
estimate of the density. Since \textit{Planck} has published a
full-sky Sunyaev--Zeldovich (SZ) catalogue, for instance, this
could be used to derive a full-sky peculiar velocity field for
galaxy clusters and compared with large-scale maps of the galaxy
distribution, refining the constraints presented here. In
addition, if the bulk motion and density distribution of
large-scale neutral hydrogen gas can be observed from future 21cm
surveys (such as \cite{ska}), our method can also be applied to
investigate non-Gaussianity in a higher redshift regime and
therefore deeper cosmic volume.


\section*{Acknowledgements}
We would like to thank Niayesh Afshordi, Neal Dalal and Ashley J.
Ross for helpful discussion, and Enzo Branchini for sharing the
PSC$z$ catalogue. YZM is supported by a CITA National Fellowship.
Part of the research is supported by the Natural Science and
Engineering Research Council of Canada.



\begin{thebibliography}{}
%
%
%




\bibitem[Bardeen et al.(1986)]{Bardeen86} Bardeen, J.~M., Bond,
J.~R., Kaiser, N., \& Szalay, A.~S.\ 1986, ApJ, 304, 15

\bibitem[\protect\citeauthoryear{Bennett et al.} {2013}]{Bennett12}
Bennett C.L. et al., 2013, ApJS, 208, 20

\bibitem[\protect\citeauthoryear{Branchini et al.} {1999}]{Branchini99} Branchini E., et al.,
1999, MNRAS, 308, 1

\bibitem[\protect\citeauthoryear{Branchini et al.} {2001}]{Branchini01} Branchini E., 2001,
MNRAS, 326, 1191
%
%
%
%
%
%
%

\bibitem[\protect\citeauthoryear{Dalal et al.} {2008}]{Dalal08} Dalal, N., Dor{\'e}, O.,
Huterer, D., \& Shirokov, A.\ 2008, Phys. Rev. D, 77, 123514

\bibitem[\protect\citeauthoryear{Davis et al.} {1996}]{Davis96} Davis M., Nusser A.; Willick J. A., 1996, ApJ,
473, 22
%
%
%
%
%
\bibitem[\protect\citeauthoryear{Folatelli et al.} {2010}]{Folatelli10} Folatelli G. et al.,
2010, AJ, 139, 120
%
\bibitem[\protect\citeauthoryear{Feldman et al.} {2001}]{Feldman01} Feldman H. A., Frieman J. A.,
Fry J. N., Scoccimarro R., 2001, Phys. Rev. Lett., 86, 1434
%
%
\bibitem[\protect\citeauthoryear{Feldman et al.} {2010}]{Feldman10} Feldman H., Watkins R., Hudson M.
J., 2010, MNRAS, 407, 2328

\bibitem[\protect\citeauthoryear{Fisher et al.} {1995}]{Fisher1995} Fisher K., Huchra J., Strauss M.,
Davis M., Yahil A., Schlegel D., 1995, ApJ, 100, 69
%
%
%
%
%
%

\bibitem[\protect\citeauthoryear{Hicken et al.} {2009}]{Hicken09} Hicken M. et al., 2009, ApJ, 700, 1097
%
%
%
%
%
%
%
%
%
%
%
%
\bibitem[\protect\citeauthoryear{Jha et al.} {2007}]{Jha07} Jha S., Riess A. G., Kirshner R. P., 2007, ApJ, 659, 122
%
%



\bibitem[Komatsu et al.(2011)]{Komatsu11} Komatsu E. et al.\ 2011, ApJS, 192,
18


\bibitem[\protect\citeauthoryear{Lewis, Challinor \& Lasenby} {2000}]{Lewis00} Lewis A., Challinor A., \& Lasenby A., 2000, ApJ, 538, 473


%
%
%
%
%

\bibitem[\protect\citeauthoryear{Ma, Gordon \& Feldman} {2011}]{Ma11}
Ma Y.Z., Gordon C., \& Feldman H., 2011, Phys. Rev. D 83, 103002.

\bibitem[\protect\citeauthoryear{Ma, Branchini \& Scott} {2012}]{Ma12a}
Ma Y.Z., Branchini E., \& Scott D., 2012, MNRAS, 425, 2880

\bibitem[Matarrese et al.(2000)]{Materrese00} Matarrese, S., Verde,
L., \& Jimenez, R.\ 2000, ApJL, 541, 10

\bibitem[Matarrese \& Verde(2008)]{Matt08} Matarrese, S., \& Verde, L.\ 2008, ApJL,
677, L77

%
%
%
%
%


\bibitem[Nikoloudakis et al.(2013)]{Nik13} Nikoloudakis, N.,
Shanks, T., \& Sawangwit, U.\ 2013, MNRAS, 429, 2032

%
%
%
%
%
\bibitem[\protect\citeauthoryear{Peebles} {1980}]{Peebles80} Peebles P. J.~E., \textit{The Large-Scale Structure of the Universe},
Princeton Univ. Press (1980).

\bibitem[\protect\citeauthoryear{Planck Collaboration XVI} {2013}]{Planck16} Planck 2013 results XVI, arXiv: 1303.5076
[astro-ph.CO].

\bibitem[\protect\citeauthoryear{Planck Collaboration XXIV} {2013}]{Planck24} Planck 2013 results XXIV, arXiv: 1303.5084 [astro-ph.CO].

%
%
\bibitem[\protect\citeauthoryear{Radburn-Smith et al.} {2004}]{Radburn-Smith04} Radburn-Smith
D.J., Lucey J.R., Hudon M.J., 2004, MNRAS, 355, 1378
%

\bibitem[\protect\citeauthoryear{Ross et al.} {2013}]{Ross13} Ross
A. et al., 2013, MNRAS, 428, 1116

%
%
\bibitem[\protect\citeauthoryear{Saunders et al.} {2000}]{Saunders00} Saunders W., et al.,
2000, MNRAS, 317, 55
%
%
%
%
\bibitem[\protect\citeauthoryear{Scoccimarro et al.} {2001}]{Scoccimarro01} Scoccimarro R.,
Feldman H. A., Fry J. N., Frieman J. A., 2001, ApJ, 546, 652
%
\bibitem[\protect\citeauthoryear{Scott \& Smoot}{2010}]{Scott10}
Scott D., Smoot G., Rev. Part. Phys. 2010, (arXiv: 1005.0555)
%
%
%
\bibitem[\protect\citeauthoryear{Square Kilometre Array} {}]{ska}
Square Kilometre Array: http://www.skatelescope.org
%
\bibitem[\protect\citeauthoryear{Springob et al.} {2007}]{Springob07} Springob C. M.,
Masters K. L., Haynes M. P., Giovanelli R., Marinoni C., 2007,
ApJS, 172, 599
%
%
%
%
%
%
%
%

\bibitem[\protect\citeauthoryear{Tully \& Fisher} {1977}]{Tully77} Tully R. B., \& Fisher J. R., 1977
A\&A, 54, 661

\bibitem[\protect\citeauthoryear{Turnbull et al.} {2012}]{Turnbull12} Turnbull S. J., Hudson
M. J., Feldman H. A., Hicken M., Kirshner R. P., Watkins R., 2012,
MNRAS, 420, 447

\bibitem[\protect\citeauthoryear{Verde et al.} {2002}]{Verde02} Verde L. et al., 2002, MNRAS, 335, 432


\bibitem[\protect\citeauthoryear{Wands \& Slosar} {2009}]{Wands09} Wands D., \& Slosar A., 2009, Phys. Rev. D, 79, 123507

\bibitem[\protect\citeauthoryear{Wands} {2010}]{Wands10} Wands D., 2010, CQG, 27, 124002


\bibitem[\protect\citeauthoryear{Wang} {2007}]{Wang07} Wang L., ApJ submitted (arXiv:
0705.0368)

\bibitem[\protect\citeauthoryear{Watkins et al.} {2009}]{Watkins09} Watkins R., Feldman H. A., Hudson
M. J., 2009, MNRAS, 392, 743
%
%
%
%


\bibitem[\protect\citeauthoryear{Xia et al.} {2011}]{Xia11} Xia J. Q., Baccigalupi C., Matarrese S., Verde L., \&
Viel M., 2011, JCAP, 08, 033

\bibitem[\protect\citeauthoryear{Yahil et al.} {1991}]{Yahil91} Yahil A., Strauss M.A., Davis M., Huchra J.P., 1991, ApJ, 372, 380


\bibitem[York et al.(2000)]{York00} York, D.G., et al. 2000, AJ, 120, 1579




\end{thebibliography}

\end{document}